\documentclass[prd,nofootinbib,letterpaper,floatfix,superscriptaddress,preprintnumbers]{revtex4}

\usepackage{color}

\usepackage{graphicx}

\usepackage{verbatim}
\usepackage{amsmath}
\usepackage{amssymb}
\usepackage{subfigure}
\usepackage{url}

\usepackage{multirow}
\usepackage{threeparttable}
\usepackage{paralist}

\usepackage{color}
\definecolor{darkgreen}{rgb}{0,0.5,0}

\newcommand{\GeV}{\text{~GeV}}
\newcommand{\TeV}{\text{~TeV}}
\newcommand{\SU}{\text{SU}}
\newcommand{\SM}{\text{SM}}

\newcommand{\U}{\text{U}}
\newcommand{\CKM}{\text{CKM}}

\newcommand{\vev}[1]{\langle #1 \rangle}

\DeclareRobustCommand{\Sec}[1]{Sec.~\ref{#1}}

\DeclareRobustCommand{\App}[1]{App.~\ref{#1}}
\DeclareRobustCommand{\Tab}[1]{Table~\ref{#1}}

\DeclareRobustCommand{\Fig}[1]{Fig.~\ref{#1}}

\DeclareRobustCommand{\Eq}[1]{Eq.~(\ref{#1})}

\DeclareRobustCommand{\Ref}[1]{Ref.~\cite{#1}}

\newcommand{\be}{\begin{equation}}
\newcommand{\ee}{\end{equation}}

\newcommand{\mb}[1]{\boldsymbol{#1}}

\begin{document}

\title{Flavor Mediation Delivers Natural SUSY}
 
\author{Nathaniel Craig}
\email{ncraig@ias.edu}
\affiliation{Institute for Advanced Study, Princeton, NJ 08540, USA \\
Department of Physics and Astronomy, Rutgers University, Piscataway, NJ 08854, USA}

\author{Matthew McCullough}
\email{mccull@mit.edu}
\affiliation{Center for Theoretical Physics, Massachusetts Institute of Technology, Cambridge, MA 02139, USA}

\author{Jesse Thaler}
\email{jthaler@mit.edu}
\affiliation{Center for Theoretical Physics, Massachusetts Institute of Technology, Cambridge, MA 02139, USA}

\date{\today}

\begin{abstract}
If supersymmetry (SUSY) solves the hierarchy problem, then naturalness considerations coupled with recent LHC bounds require non-trivial superpartner flavor structures.  Such ``Natural SUSY'' models exhibit a large mass hierarchy between scalars of the third and first two generations as well as degeneracy (or alignment) among the first two generations.  In this work, we show how this specific beyond the standard model (SM) flavor structure can be tied directly to SM flavor via ``Flavor Mediation''.  The SM contains an anomaly-free $\SU(3)$ flavor symmetry, broken only by Yukawa couplings.  By gauging this flavor symmetry in addition to SM gauge symmetries, we can mediate SUSY breaking via (Higgsed) gauge mediation.  This automatically delivers a natural SUSY spectrum.   Third-generation scalar masses are suppressed due to the dominant breaking of the flavor gauge symmetry in the top direction.  More subtly, the first-two-generation scalars remain highly degenerate due to a custodial $\U(2)$ symmetry, where the $\SU(2)$ factor arises because $\SU(3)$ is rank two.  
 This custodial symmetry is broken only at order $(m_c/m_t)^2$.  SUSY gauge coupling unification predictions are preserved, since no new charged matter is introduced, the SM gauge structure is unaltered,  and the flavor symmetry treats all matter multiplets equally.  Moreover, the uniqueness of the anomaly-free $\SU(3)$ flavor group makes possible a number of concrete predictions for the superpartner spectrum.  
\end{abstract}

\preprint{MIT-CTP {4350}, RU-NHETC {2012-03}}

\maketitle

\section{Introduction}
\label{sec:introduction}

Recent LHC searches for weak-scale supersymmetry (SUSY) have emphasized the importance of flavor structures for physics beyond the standard model (SM).  Direct searches place stringent bounds on first-generation squarks, pushing their masses above around 1 TeV \cite{Chatrchyan:2011zy,Aad:2011ib}.  If all squark flavors were degenerate,  however, this would imply some degree of fine-tuning in the Higgs potential, since third-generation squarks must be closer to 500 GeV to satisfy naturalness considerations (see, e.g., \cite{Kitano:2006gv} and references therein).   This tension has renewed interest in models of ``Natural SUSY'' \cite{Dimopoulos:1995mi,Dine:1993np,Pouliot:1993zm,Barbieri:1995rs,Pomarol:1995xc,Barbieri:1995uv,Cohen:1996vb,Dvali:1996rj,Barbieri:1997tu,Kaplan:1998jk,Kaplan:1999iq,Gabella:2007cp,Sundrum:2009gv,Barbieri:2010pd,Barbieri:2010ar,Craig:2011yk,Gherghetta:2011wc,Jeong:2011en,Essig:2011qg,Kats:2011qh,Papucci:2011wy,Brust:2011tb,Delgado:2011kr,Desai:2011th,Akula:2011jx,Ajaib:2011hs,Ishiwata:2011fu,Lodone:2011pv,He:2011tp,Arvanitaki:2011ck,Auzzi:2011eu,Csaki:2012fh,Craig:2012yd, Larsen:2012rq,Craig:2012hc} which achieve the desired splitting between the squark generations to satisfy LHC bounds while still addressing the (little) hierarchy problem.

An important requirement for natural SUSY models is flavor degeneracy or alignment among the squarks of the first two generations.  In particular, natural SUSY models often invoke an approximate $\U(2)$ symmetry between the first two generations in order to avoid otherwise large flavor-changing processes.  In such models, flavor beyond the SM has a very different structure from flavor within the SM, and one therefore seeks an explanation for how such flavor structures might arise in a realistic theory.    

The goal of this paper is to show how SUSY flavor can be tied directly to SM flavor via ``Flavor Mediation''.\footnote{This mechanism is somewhat in the spirit of \Ref{Kaplan:1998jk,Kaplan:1999iq}, where SUSY breaking was communicated via a Froggatt-Nielsen $\U(1)$.  It should not be confused with flavored gauge medation \cite{Shadmi:2011hs}, where messengers have tree-level couplings to the visible sector.  For recent constructions involving gauged flavor symmetries see \Ref{Grinstein:2010ve,Guadagnoli:2011id}.}   We gauge an anomaly-free $\SU(3)_F$ subgroup of the SM flavor symmetries that is then spontaneously broken to yield SM Yukawa couplings.  By charging SUSY-breaking messengers under $\SU(3)_F$, we generate SM sfermion masses via ``Higgsed gauge mediation'' \cite{Gorbatov:2008qa,Craig:2012yd}.  Intriguingly, if the flavor-breaking scale and the messenger scale are close, then the third-generation squarks are significantly lighter than the first- and second-generation squarks.  Moreover, the first- and second-generation squarks are automatically highly degenerate, satisfying flavor bounds without having to impose a $\U(2)$ symmetry by hand.

\begin{figure}[t]
\centering
\includegraphics[height=1.3in]{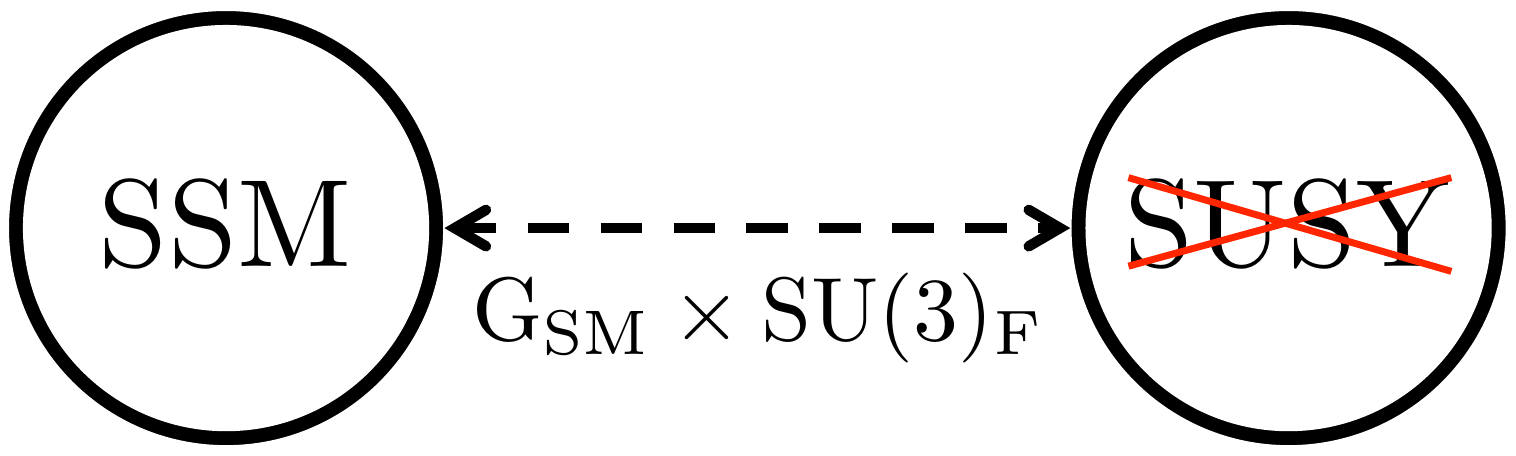}
\caption{A depiction of flavor mediation where SUSY breaking is communicated to the SSM by both SM and flavor gauge groups.  SUSY breaking in a hidden sector is communicated by messenger superfields at one loop to the $G_{\SM} \times \SU(3)_F \equiv \SU(3)_C \times \SU(2)_L \times \U(1)_Y \times \SU(3)_F$ gauge superfields, and at two loops to the SSM chiral superfields charged under these symmetries.  This generates standard gauge-mediated soft masses for the SM gauginos and approximately diagonal soft masses for all SSM scalars.   Sfermions of the first two generations obtain large, degenerate soft masses from flavor mediation with small, generation-independent splittings due to gauge mediation from the SM gauge groups.  Third-generation sfermions obtain comparable soft mass contributions from all gauge groups.}
\label{fig:flavgaugediag}
\end{figure}

The key to achieving a realistic SUSY flavor structure is in the fact that $\SU(3)_F$ is rank two \cite{Craig:2012yd}.  When the gauged flavor group is spontaneously broken by the large top Yukawa coupling, there is a residual approximate custodial $\U(2)$ symmetry, of which the $\SU(2)$ subgroup is gauged, which shields first-two-generation scalars from the hierarchy in first-two-generation Yukawas.   In this way, flavor mediation can deliver all the desired features of natural SUSY.

A complete model of flavor mediation is shown in \Fig{fig:flavgaugediag}, where both the flavor gauge group and SM gauge groups participate in (Higgsed) gauge mediation to the supersymmetric standard model (SSM).  Since the SM Higgs multiplets do not carry flavor quantum numbers, they are naturally lighter than the flavored sfermions, as needed to minimize fine-tuning.  Since SM gauginos only get their masses from SM gauge mediation, they are also typically light.   After accounting for renormalization group (RG) effects, the gluinos end up being a bit heavier than the third-generation squarks, perfect for a natural SUSY spectrum.

The uniqueness of the anomaly-free $\SU(3)_F$ leads to a number of interesting predictions.  First, because the flavor gauge group is broken by SM Yukawa matrices, the hierarchy between the third-generation squarks and the first- and second-generation squarks cannot be made arbitrarily large.  Thus, a discovery of light stops and sbottoms would yield an upper bound for the masses of the remaining squarks.  Second, in order for $\SU(3)_F$ to be anomaly-free, both leptons and quarks must be charged under the flavor symmetry, so one expects light staus and third-generation sneutrinos to be accessible at LHC energies.  Third, while generic natural SUSY models do not require a right-handed sbottom in the spectrum, flavor mediation treats right-handed stops and sbottoms democratically, with the only splitting arising from SM gauge mediation and RG effects.  Finally, flavor mediation preserves many of the desired features of SUSY grand unified theories (GUTs).  Since the anomaly-free $\SU(3)_F$ does not require any new SM-charged chiral matter and treats all matter multiplets equally, SUSY gauge coupling unification is preserved.  Assuming gauge mediation is the dominant source for gaugino masses, then SM gaugino masses also unify.

The outline for the remainder of this paper is as follows.  In \Sec{sec:model}, we introduce the anomaly-free $\SU(3)_F$ flavor gauge group and describe how it is broken.  In \Sec{sec:sfermions}, we describe the physics of flavor mediation, and how the massive flavor gauge bosons contribute to the sfermion spectra via Higgsed gauge mediation.  We outline a complete model in \Sec{sec:completemodel}, detailing the generation of gaugino masses in \Sec{sec:gauginos}, the Higgs sector in \Sec{sec:higgs}, and typical sparticle spectra in \Sec{sec:exampspectra}.  We verify in \Sec{sec:constraints} that flavor bounds are satisfied in this model.  We sketch the key predictions of our model in \Sec{sec:predictions} and conclude in \Sec{sec:conclusion}.

\section{The Gauged Flavor Symmetry}
\label{sec:model}

\subsection{Motivating $\SU(3)_F$}

A wide range of flavor symmetries have been proposed to explain some or all features of the quark and lepton mass matrices and mixings.  As our goal is to link SM flavor structures with a natural SUSY soft mass spectrum, we must employ some additional guiding (or at least simplifying) principles to select a preferred gauged flavor symmetry.

First, the flavor symmetry should act equally on all three generations.  There are SUSY models employing additional gauged $\U(1)$, $\SU(2)$, or $\U(2)$ flavor symmetries that can achieve a natural SUSY spectrum \cite{Dine:1993np,Pouliot:1993zm,Pomarol:1995xc,Cohen:1996vb,Dvali:1996rj,Kaplan:1998jk,Kaplan:1999iq,Delgado:2011kr}.  However, it is somewhat ad hoc to treat the first two generations separately from the third without some underlying reason.  By treating all generations on an equal footing, one can more easily obtain the SM mass and mixing structure.

Second, the flavor symmetry should act equally on lepton and quark multiplets in order to allow for a GUT structure in the ultraviolet (UV).  This is further motivation to treat all three generations equally, since $\U(1)$, $\SU(2)$, or $\U(2)$ flavor symmetries make it difficult to explain the near maximal neutrino mixing between the second and third generations.  Of course, one can always imagine split GUT multiplets where quarks and leptons do not live in the same GUT multiplet, but we choose not to consider that possibility.

Third and finally, we wish to avoid adding additional chiral matter with SM gauge charges in the infrared (IR), in order to maintain SUSY gauge coupling unification in the UV.   Many candidate flavor symmetries, particularly ones involving $\U(1)$s, are anomalous, requiring the addition of further matter with SM charges to cancel the anomalies. We can avoid extra charged multiplets if the flavor symmetry has no SM gauge anomalies.  In essence, this corresponds to taking the MSSM in the limit of zero Yukawas and gauging any anomaly-free symmetry compatible with GUT structures.\footnote{If we add right-handed neutrinos, we could also gauge $\U(1)_{B-L}$ consistent with this philosophy.} 

\begin{table}[t]
\centering
\begin{tabular}{c | c c c c c c c | c c c}
\hline\hline
 & $\mb{Q}$ & $\mb{U^c}$ & $\mb{D^c}$ & $\mb{L}$& $\mb{E^c}$&$\mb{H_u}$& $\mb{H_d}$& $\mb{N^c}$&  $\mb{S_u}$& $\mb{S_d}$ \\ [0.5ex]  \hline

  $\SU(3)_F$ & $\mb{3}$ &  $\mb{3}$ &  $\mb{3}$ &  $\mb{3}$ &  $\mb{3}$ &    $\mb{1}$ &  $\mb{1}$ & $\mb{\bar 3}$ &  $\mb{\bar 6}$ &  $\mb{\bar 6}$  \\
\hline \hline
\end{tabular}
\caption{The $\SU(3)_F$ charges of the minimal  chiral matter representations required for anomaly-free flavor mediation. The fields $\mb{N^c}, \mb{S_u}, \mb{S_d}$ are all neutral under $G_{\SM}$; their role will be discussed in \Sec{subsec:yukawas}. Additional pairs of massive vector-like representations may be included to generate Yukawa interactions, mediate supersymmetry breaking, and generate lepton flavor structures.}
\label{tab:matter}
\end{table}

Fortunately, there is a well-known flavor symmetry satisfying all of these requirements: an $\SU(3)_F$ flavor symmetry under which all SSM matter supermultiplets are fundamentals. The charge assignments of SSM supermultiplets under $\SU(3)_F$ are shown in \Tab{tab:matter}.  In fact, this is the maximal group involving $\SU(3)$ factors that is anomaly-free and treats all matter multiplets equally.\footnote{Extending this group to $\U(3)$ is not possible as the additional $\U(1)$ factor is anomalous.  As will be discussed later, the fact that we are forced to employ an $\SU(3)$, rather than $\U(3)$, symmetry is very appealing for the generation of a natural SUSY spectrum through flavor mediation.  A remaining $\U(1)$ factor would generate additional, undesirable, splittings between first- and second-generation squarks.}  A number of successful flavor models employing this symmetry have been constructed \cite{Chkareuli:1981gy,Berezhiani:1982rr,Berezhiani:1983hm,Berezhiani:1985in,Anselm:1985we,Berezhiani:1996ii,Soldate:1986rq,Koide:1987vp,Berezhiani:2000cg,Kitano:2000xk,King:2001uz,King:2003rf}, including models that allow for GUT multiplets in the UV.  

\subsection{Yukawa Couplings}\label{subsec:yukawas}

In order to generate SM Yukawa couplings, the flavor group must be spontaneously broken.  Clearly, the Yukawas must transform as a $\mb{\overline{3}} \otimes \mb{\overline{3}}$ under the $\SU(3)_F$ symmetry.  They could arise as the sum of pairs of fundamental representations, in which case the SM Yukawa coupling  will be generated through a dimension-six operator and will depend on the square of vacuum expectation values (vevs).  Alternatively, the Yukawas could arise from a dimension-five operator through a symmetric or antisymmetric two-index representation.  

As we will see, the effects of flavor mediation are enhanced by having a large hierarchy between the flavor boson masses, which is desirable to achieve a natural SUSY spectrum.  With pairs of fundamentals, the flavor boson masses will be parametrically proportional to the square root of the SM Yukawas.  With a two-index representation, the flavor boson masses will be linear in the SM Yukawas.  Therefore, in order to generate the largest possible mass hierarchy, we will employ two-index representations.  This also comes with the advantage of requiring fewer messenger superfields in order to generate the Yukawas.  A single antisymmetric representation does not have sufficient rank to generate the SM Yukawas, so from now on we will consider symmetric representations.

First considering just quark superfields, we add two symmetric $\mb{\overline{6}}$ representations of $\SU(3)_F$ to the SSM, which we denote as $\mb{S_{u,d}}$. The cubic $\SU(3)_F$ anomalies vanish if we also add right-handed neutrino superfields $\mb{N^c}$ transforming as a $\mb{\overline{3}}$.\footnote{This renders $\SU(3)_F$ IR-free.  If the Landau pole in the gauge coupling lies below the scale of UV physics such as the GUT or Planck scale, a dual description would be required in order to UV-complete the model. However, the scale of $\SU(3)_F$ breaking may readily lie at or above the GUT scale, thereby avoiding any Landau poles beneath the gauge cutoff.}  The SM quark Yukawas can be generated through the higher-dimensional superpotential operators
\be
W = \frac{1}{M_{S_u}} \mb{S_u H_u Q U^c} + \frac{1}{M_{S_d}} \mb{S_d H_d Q D^c} ,
\ee
where flavor indices have been suppressed.  These operators can arise by integrating out heavy vector-like Higgs pairs also in the $\mb{6,\overline{6}}$ of $\SU(3)_F$ with mass $M_S$; unification is preserved in the usual way if these Higgses live in complete multiplets of $\SU(5)_{\rm GUT}$. In particular, all SM quark Yukawas may be generated by integrating out fields transforming as $(\mb{\bar 5, 6}) \oplus (\mb{5, \bar 6}) \oplus (\mb{5, 6}) \oplus (\mb{\bar 5, \bar 6})$ under $\SU(5)_{\rm GUT} \times \SU(3)_F$. This suggests the scale $M_S$ should be high enough to avoid inducing Landau poles in the SM gauge couplings below the unification scale, and an $\mathcal{O}(1)$ top Yukawa then implies that the flavor symmetry breaking scale should also be high.

One can also introduce additional anomaly-free representations of $\SU(3)_F$ in order to generate charged lepton Yukawas and neutrino masses.  However, without committing to a particular model of neutrino mass generation, the less constrained leptonic flavor structure means that it is difficult to extract general features of the role lepton masses and mixings play in the breaking of $\SU(3)_F$, and subsequently the flavor-mediated soft masses.  Henceforth, we will assume that the dominant breaking of the flavor symmetry lies in the generation of the quark Yukawas.  We will see in \Sec{subsec:mixings} that this is in fact a good approximation, and so for now it suffices to only consider the quark sector.

\subsection{Flavor Breaking}
To have a realistic model, the vevs of $\mb{S_{u,d}}$ must generate the SM quark flavor structures. As the goal of this work is to connect the SM fermion flavor structure to sfermion flavor structure, and numerous patterns of $\SU(3)_F$ breaking already exist in the literature (see e.g.\ \cite{Berezhiani:2000cg}), we simply treat $\mb{S_{u,d}}$ as flavor spurions which obtain vevs in a supersymmetric manor along a $D$-flat direction.\footnote{The hierarchical structure of vevs could arise from higher-dimensional operators, gauge dynamics, radiative effects, or some other mechanism.  In general, one expects that in an explicit model, there will be additional matter charged under $\SU(3)_F$ beyond \Tab{tab:matter}.  The following analysis assumes that vevs of the additional fields are either proportional to $\vev{\mb{S_{u,d}}}$ or give a subdominant contribution to $\SU(3)_F$ breaking. However, the qualitative conclusions are typically not changed by the presence of additional vevs, provided the primary breaking $\SU(3)_F \rightarrow \SU(2)_F$ is still aligned with the top Yukawa and the subsequent breaking $\SU(2)_F \rightarrow \emptyset$ occurs at a parametrically lower scale.}  After performing an $\SU(3)_F$ rotation, we can assume vevs of
\be \label{eq:vevs}
\langle \mb{S_u} \rangle = \left( \begin{array}{ccc} v_{u1} & 0 & 0 \\ 0 & v_{u2} & 0 \\ 0 & 0 & v_{u3}  \end{array} \right), \qquad  \langle \mb{S_d} \rangle = V_{\CKM} \left( \begin{array}{ccc} v_{d1} & 0 & 0 \\ 0 & v_{d2} & 0 \\ 0 & 0 & v_{d3}   \end{array} \right) V_{\CKM}^T.
\ee
Here, the CKM mixing matrix $V_{\CKM}$ arises due to the initial misalignment of the two vevs, and we assume the hierarchy $v_{u3} \gg v_{u2} \gg v_{u1}$ and $v_{d3} \gg v_{d2} \gg v_{d1}$.

In this vacuum, the gauge symmetry is fully broken and SSM quark Yukawa couplings are generated.  There is some freedom in choosing the relative scales of the up and down flavor symmetry breaking vevs.  In particular, $M_{S_u}$ need not equal $M_{S_d}$, and there is freedom in the Yukawas through the ratio of the up- and down-type Higgs vevs $\tan \beta \equiv \vev{H_u}/\vev{H_d}$.  We can parameterize both freedoms with one parameter $\alpha$, since
\be
\label{eq:defalpha}
\frac{m_t}{m_b} = \frac{v_{u3}}{v_{d3}} \alpha,   \qquad \alpha \equiv \frac{M_{S_d}}{M_{S_u}} \tan \beta .
\ee
Varying $\alpha$ then leads to different flavor boson spectra.  

\begin{figure}[t]
\centering
\includegraphics[height=3.0in]{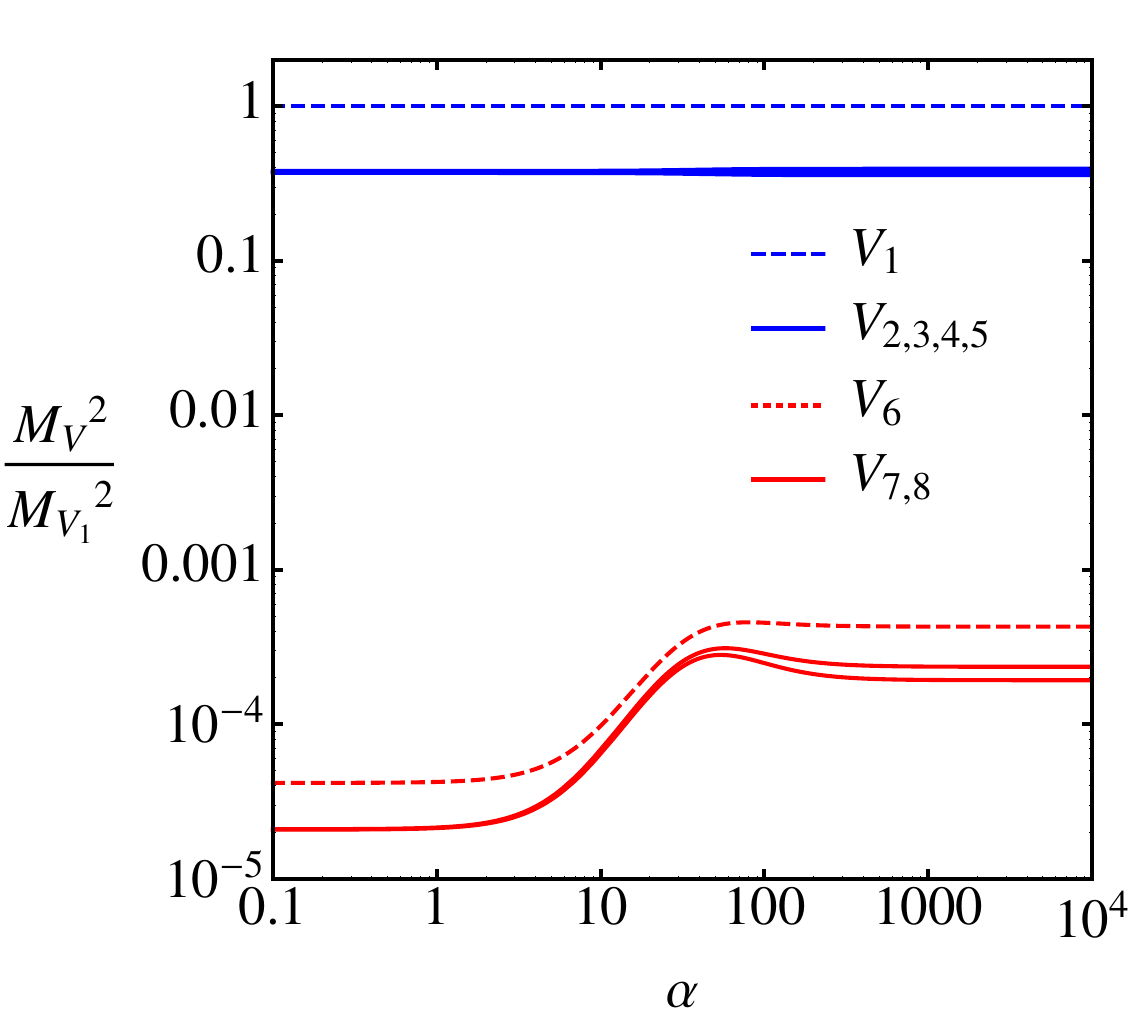}
\caption{The spectrum of flavor bosons as a function of $\alpha =  (M_{S_d}/M_{S_u}) \tan \beta$.  The masses are plotted relative to the mass of the heaviest boson, denoted by $V_1$.  The five masses at the top of the plot corresponding roughly to the $\SU(3)_F/\SU(2)_F$ generators.  The three masses at the bottom corresponding to the remaining $\SU(2)_F$ symmetry broken at much smaller scales.  The hierarchy is greatest in the limit of small $\alpha$, where the breaking is dominated by the up-quark Yukawas.  Furthermore, the two lightest bosons are more degenerate in the small $\alpha$ limit since $m_c/m_t < m_s/m_b$.  Thus, $\alpha \lesssim 100$ is favored for a natural SUSY spectrum.}
\label{fig:flavspec}
\end{figure}

For small $\alpha$ the breaking of $\SU(3)_F \rightarrow \SU(2)_F \rightarrow \emptyset$ is determined dominantly by the hierarchies in the up-quark mass matrix, whereas for large $\alpha$ the down-quark mass matrix dominates the breaking pattern.  In \Fig{fig:flavspec}, we plot how the relative spectrum of flavor boson masses varies as a function of $\alpha$.  To achieve the largest hierarchy in the sfermion masses, $\alpha \lesssim 100$ is preferred for the generation of a natural SUSY soft spectrum.  For this reason, and to limit the free parameters, we choose to set $\alpha =1$ for the remainder of this paper, simply noting that other values are also valid.

To understand the flavor boson hierarchies, consider only the flavor breaking from $\mb{S_u}$.  To second order in $v_{u2}$, the spectrum of flavor bosons is 
\be \label{eq:gbmass}
M_V^2 = g_F^2  \left\{ \frac{8}{3} v_{u3}^2, (v_{u3}+v_{u2})^2,v_{u3}^2,v_{u3}^2,(v_{u3}-v_{u2})^2, 2 v_{u2}^2, v_{u2}^2, v_{u2}^2 \right\} ,
\ee
where $g_F$ is the flavor gauge coupling.  The hierarchy $v_{u3} \gg v_{u2}$ leads to a flavor gauge boson hiearchy with five heavy gauge bosons (corresponding roughly to the generators of $\SU(3)_F/\SU(2)_F$) and three light gauge bosons (the remaining $\SU(2)_F$ generators).

For simplicity, we will use the notation $v_{u3} \equiv v_F$, since the dominant breaking is by the top Yukawa coupling.  Including the down-type Yukawas and fixing $\alpha = 1$, the other parameters are now fixed by the measured fermion masses.  The relative spectrum of flavor bosons is then set, with the overall mass scale depending only on $g_F v_F$.  In descending order, the gauge boson masses are 
\be
\begin{array}{rcl}
M_V^2 [\sim \SU(3)_F/\SU(2)_F] & = &  g_F^2 v_F^2  \left\{ 2.67, 1.02, 1.00, 1.00, 0.99 \right\}, \\
& &\\
M_V^2 [\sim \SU(2)_F] & = &   g_F^2 v_F^2   \left\{  1.13, 0.57 , 0.57 \right\} \times 10^{-4}. \\
\end{array}
\ee
Again, one can clearly see the approximate symmetry breaking pattern $\SU(3)_F \rightarrow \SU(2)_F$ followed by $\SU(2)_F \rightarrow \emptyset$ where the $\SU(2)_F$ is broken a further two orders of magnitude below.  In the next section, we demonstrate why such a structure is highly appealing for the generation of a natural SUSY spectrum.

\section{Sfermion Spectrum}
\label{sec:sfermions}

In flavor mediation, the flavor gauge symmetry is used to mediate SUSY breaking to the SSM.  Because the flavor gauge group is spontaneously broken, this leads to a model of Higgsed gauge mediation \cite{Gorbatov:2008qa,Craig:2012yd}, which depends both on the messenger scale and gauge breaking scale.  Following the usual procedure for gauge-mediated scenarios, we add a pair of messenger superfields $\mb{\Phi}$/$\mb{\Phi^c}$ in a vector-like representation of $\SU(3)_F$ and charged under a messenger-parity symmetry.  We couple these fields to a SUSY breaking spurion $\mb{X}$ using the superpotential
\be
\mb{W} = \mb{X \Phi \Phi^c},
\ee
with the assumed vev $\langle \mb{X} \rangle = M + \theta^2 F$.  At two loops, any scalars charged under $\SU(3)_F$ will obtain soft masses, in particular the squarks and sleptons of the SSM.

\subsection{Mass Eigenstates}
\label{subsec:masses}

The calculation of two-loop soft mass contributions in Higgsed gauge mediation was first performed in \Ref{Gorbatov:2008qa}.   A more transparent derivation using analytic continuation into superspace was shown in \Ref{Craig:2012yd}.  This method gives results valid to lowest order in $F/M$, for any mediating gauge group with arbitrary breaking pattern, and yields compact expressions which we now review.

Once the gauge symmetry is broken, we can simultaneously diagonalize the (SUSY) gauge boson mass matrix and corresponding group generators, such that an eigenstate with mass $M_V^a$ is associated with the generator $T^a$.  After performing this diagonalization, the resulting expression for sfermion soft masses at two loops is
\be
\left(\widetilde{m}_q^2 \right)_{ij} = C(\mb{\Phi}) \frac{\alpha_F^2}{(2 \pi)^2}  \left | \frac{F}{M} \right |^2  \sum_a f(\delta^a) \, (T_q^a T_q^a)_{\{ij\}}, \qquad \delta^a \equiv \frac{{M^a_V}^2}{M^2},
\label{eq:nonab}
\ee
where $\{ij\}$ indicates that these indices have been symmetrized, $\alpha_F \equiv g_F^2/4\pi$ is the fine structure constant for the flavor gauge group, and $C(\mb{\Phi})$ is the Dynkin index of the messenger superfield representation.  The suppression factor $f(\delta^a)$ tracks the difference between Higgsed gauge mediation and ordinary gauge mediation, and is given explicitly by
\be
f(\delta) = 2 \frac{\delta (4-\delta) ((4-\delta) +(\delta+2) \log(\delta))+2 (\delta-1) \Omega(\delta)}{\delta (4-\delta)^3},
\label{eq:fd}
\ee
with
\be
\Omega (\delta) = \sqrt{\delta (\delta-4)} \left(2 \zeta (2) + \log^2 \left(\alpha \right) + 4 \text{Li}_2 \left[-\alpha \right] \right), \qquad \alpha = \left(\sqrt{\frac{\delta}{4}} +\sqrt{\frac{\delta}{4} -1} \right)^{-2}.
\label{eq:omega}
\ee
When $\delta = 0$, $f(0) = 1$ gives the results from ordinary gauge mediation.  For large $\delta$, $f(\delta) \simeq 2(\log \delta -1)/\delta$.

Applying these results to the flavor group and breaking pattern described in \Sec{sec:model}, the soft mass-squared for the $i$-th flavor of squark or slepton in a representation of $\SU(3)_F$ is
\be
\left(\widetilde{m}_q^2 \right)_{ii} = \gamma_i (\delta) C(\mb{\Phi}) C_2 (\mb{q}) \frac{\alpha_F^2}{(2 \pi)^2}  \left | \frac{F}{M} \right |^2,  \qquad  \delta = \frac{g_F^2 v_F^2}{M^2},
\label{eq:nonabFinal}
\ee
where $C_2 (\mb{q})$ is the quadratic Casimir of the quark superfield $\mb{q}$, and $\gamma_i (\delta)$ is a generation-dependent suppression factor arising due to the breaking of the mediating gauge group, with the limiting behavior
\be
\lim_{v_F\rightarrow 0} \gamma_i (\delta) = 1  .
\ee

\begin{figure}[t]
\centering
\includegraphics[height=2.8in]{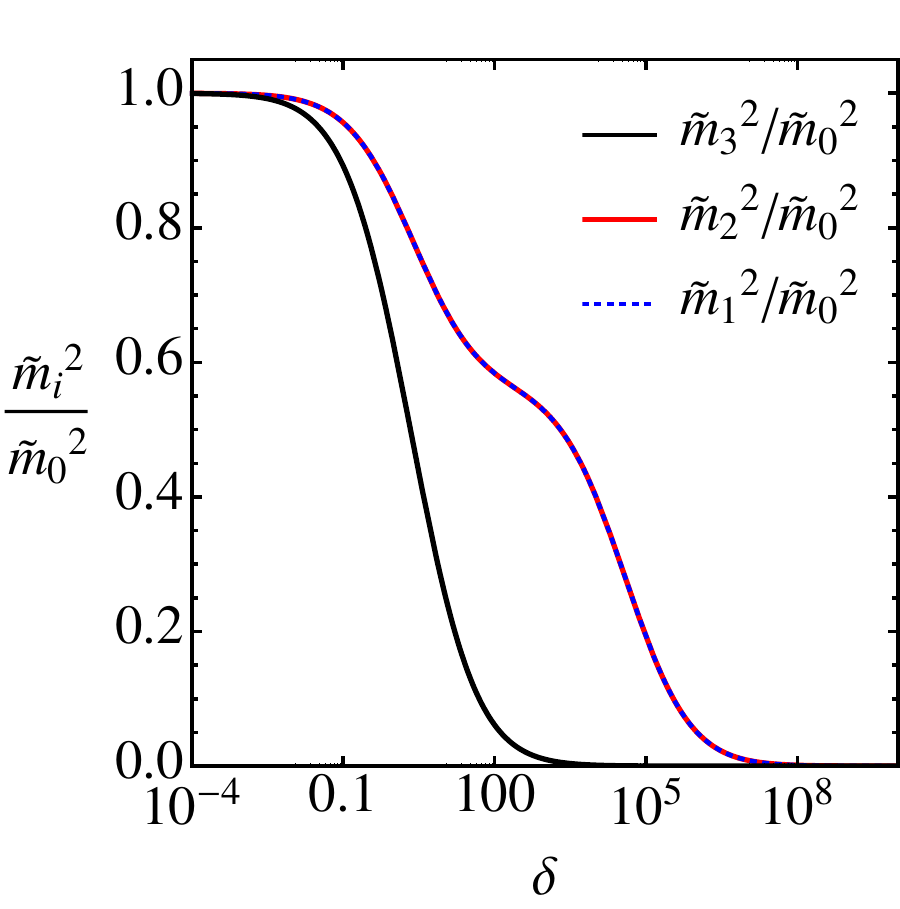}  \hspace{0.4in} \includegraphics[height=2.8in]{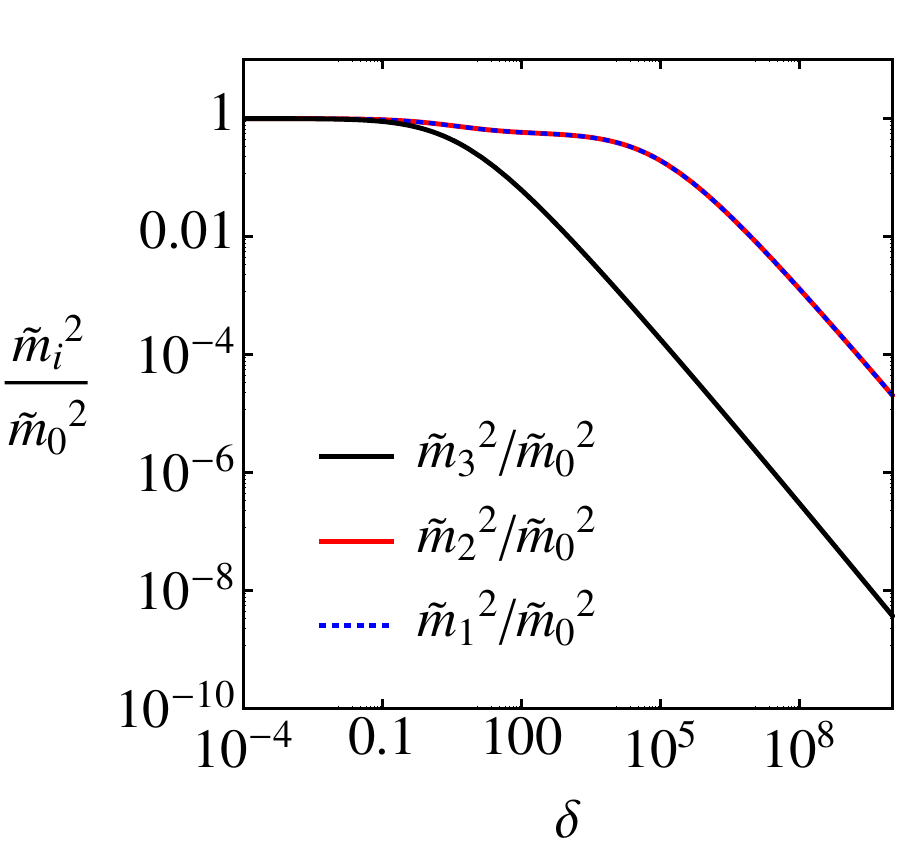}    
\caption{The $\SU(3)_F$ flavor-mediated sfermion soft mass spectrum relative to the unbroken case as a function of $\delta = g_F^2 v_F^2/M^2$.  The mass-squared is plotted linearly in the left panel, and logarithmically in the right.  There is an overall suppression which occurs whenever the flavor-breaking scale becomes comparable to the messenger scale.   The suppression of the third-generation sfermion soft masses relative to the first two generations is also clear, arising from the fact that the dominant flavor symmetry breaking lies in the top-quark direction.  The first two generations are highly degenerate, as expected from the $\SU(3)_F \rightarrow \SU(2)_F \rightarrow \emptyset$ breaking structure.}
\label{fig:sfspec}
\end{figure}

In \Fig{fig:sfspec}, we plot the suppression of the sfermion soft masses compared to the case where the gauge group is unbroken, for a range of values of $\delta$.  In the limit where the gauge group is largely unbroken, the suppression of all three soft masses is negligible.  Whenever the scale at which the gauge group is broken becomes comparable to or greater than the messenger mass scale, the suppression becomes quite significant.

\begin{figure}[t]
\centering
\includegraphics[height=2.8in]{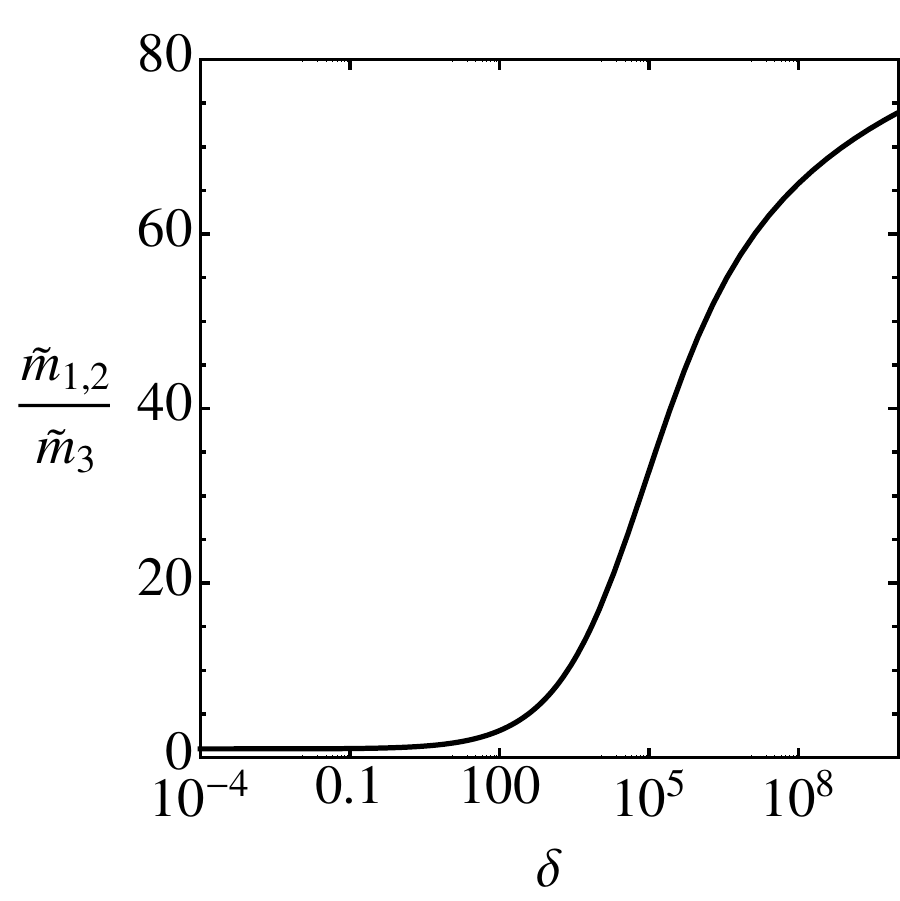}  \hspace{0.4in} \includegraphics[height=2.8in]{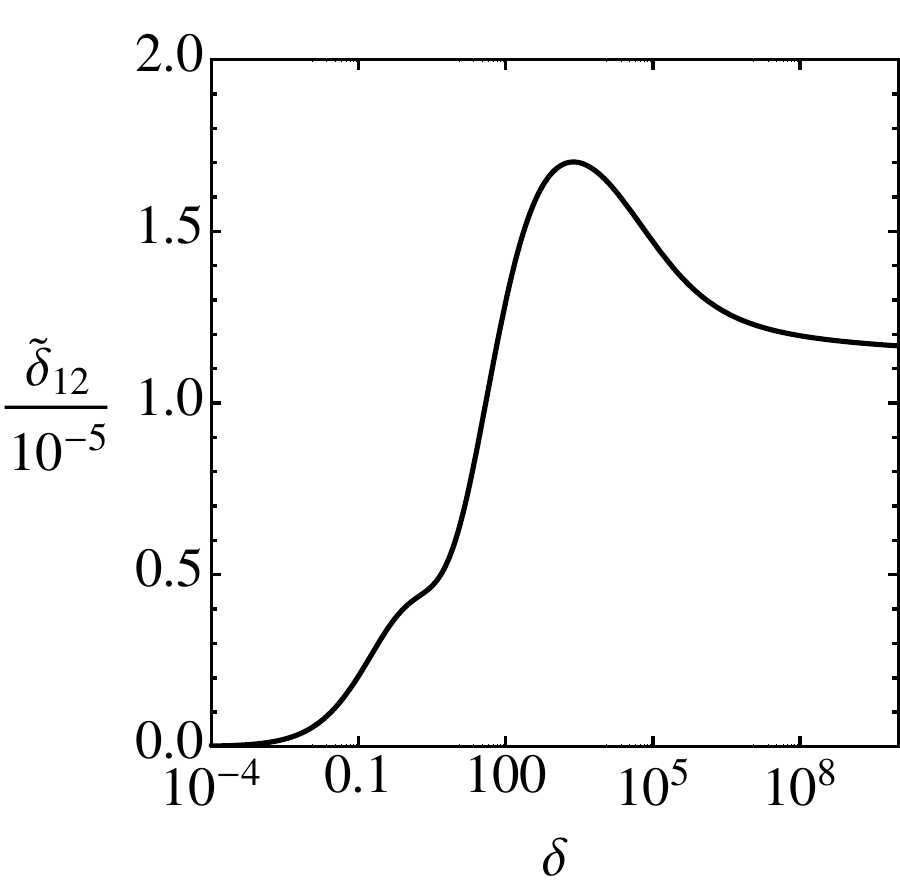}    
\caption{Intergenerational sfermion mass splittings from flavor mediation as a function of $\delta = g_F^2 v_F^2/M^2$.  In the left panel, the mass of the first-two-generation sfermions is shown relative to the third-generation.  This splitting becomes important whenever the messenger mass scale approaches the scale at which $\SU(3)_F$ is broken to an approximate $\SU(2)_F$.  The splitting endures even when the scale of $\SU(2)_F$ breaking is greater than the messenger mass scale, however it never exceeds a ratio of $\sim 100$.  In the right panel, we show the relative mass-squared splitting of the first-two-generation squarks, $\tilde{\delta}_{12} = (\tilde{m}^2_2-\tilde{m}^2_1)/\overline{m}^2$, where $\overline{m}^2 = (\tilde{m}^2_2+\tilde{m}^2_1)/2$.  This splitting remains below $2 \times 10^{-5}$ regardless of the relative scales of flavor breaking and messenger masses.}
\label{fig:sfsplit}
\end{figure}

In \Fig{fig:sfsplit}, we plot the various splittings between sfermion generations.  When the gauge group is largely unbroken, the third generation is almost degenerate with the first two.  As the breaking of $\SU(3)_F \rightarrow \SU(2)_F$ becomes significant, a large splitting between the third- and first-two-generation sfermion masses emerges.  This splitting remains even when the breaking of the remaining $\SU(2)_F$ becomes comparable to the messenger scale, although it never exceeds a ratio of greater than $\sim 100$.  The splitting between the first-two-generation sfermions is also shown, which first appears at order $(m_c/m_t)^2$.  Flavor-changing neutral current processes are very sensitive to this splitting, and hence it is important for a successful model that this splitting is small.  Regardless of the relative hierarchy between the flavor-breaking scale and the messenger scale, the squared-mass splitting never exceeds a fractional value of $2 \times 10^{-5}$.

The soft mass spectrum generated via $\SU(3)_F$ flavor mediation is extremely attractive from the perspective of natural SUSY.   Relative mass splittings of $\mathcal{O}(10)$ between the third and first two generations can be easily accommodated, while mass splittings between the first two generations remain very small.  All three generations are treated on a equal footing, and the mass splittings arise solely due to the flavor-symmetry breaking implied by the quark masses.  The fact that only an $\SU(3)$, rather than $\U(3)$, symmetry is anomaly free means that the symmetry breaking structure is $\SU(3)_F \rightarrow \SU(2)_F \rightarrow \emptyset$, protecting the first two generations from mass splittings.  Were it possible to gauge a $\U(3)$ symmetry, the remaining $\U(1)$ in the symmetry breaking structure $\U(3)_F \rightarrow \U(2)_F \rightarrow \U(1)_F \rightarrow \emptyset$ would have generated additional splittings between the first two generations.

\subsection{Mixing Angles}
\label{subsec:mixings}

An interesting feature of flavor mediation is that phases and mixing angles from the SM CKM matrix are transmitted to the scalar soft mass matrix via the gauged flavor symmetry.  In the model presented, phases and mixing angles in the symmetry breaking vevs generate mixings in the flavor boson mass matrix.   These then show up in the scalar soft mass matrices through \Eq{eq:nonab}, since gauge boson mass eigenstates have off-diagonal elements in the generator basis.  

The full soft mass matrix is a non-trivial function of the vevs and angles in \Eq{eq:vevs}, and includes a dependence on the function $f(\delta^a)$.  To gain insight into the magnitude of these terms, we will perform a perturbative calculation valid when mixing angles are relatively small and the $\SU(2)_F$ gauge bosons are much lighter than the $\SU(3)_F/\SU(2)_F$ gauge bosons, i.e. $\max [ v_{u3}, v_{d3} ] \gg \max [ v_{u2}, v_{d2} ] $.  For the case at hand, it is a good approximation to work in the limit of vanishing $ v_{u1}, v_{d1}$.   To first order in the mixing angles and zeroth order in $v_{u2}/\sqrt{v_{u3}^2+v_{d3}^2}$ and $v_{d2}/\sqrt{v_{u3}^2+v_{d3}^2}$, the magnitudes of the resulting soft mass matrix elements are
\be \label{eq:softmixings}
 \tilde{m}^2  \approx \left( \begin{array}{ccc} \tilde{m}^2_2 & 0 &0 \\ 0 & \tilde{m}^2_2 & 0 \\ 0 & 0 & \tilde{m}^2_3  \end{array} \right)  +
 (\tilde{m}^2_2-\tilde{m}^2_3) \frac{v_{d3}^2}{v_{u3}^2+v_{d3}^2} \left( \begin{array}{ccc} 0 & 0 & \cos (\delta_{\CKM}) |V_{13}| \\ 0 & 0 & |V_{23}| \\ \cos (\delta_{\CKM}) |V_{13}| & |V_{23}| & 0  \end{array} \right)  ,
\ee
where the approximate degeneracy of the first-two-generation scalar masses has been taken account of in the diagonal components.  

There are a number of interesting features of \Eq{eq:softmixings}.  While $V_{12}$ is the largest mixing in the CKM matrix by a considerable amount, terms proprotional to $V_{12}$ are absent from our expansion since they come suppressed by a factor of $v_{d2}^2/(v_{u3}^2+v_{d3}^2)$ which is always small.  In the limit $v_{d3} \ll v_{u3}$, the off-diagonal components are suppressed and the soft masses become approximately diagonal.  In the limit $v_{d3} \gg v_{u3}$ where the flavor-symmetry breaking is driven dominantly by the down sector, the off-diagonal components become large and correspond to a rotation determined by the CKM matrix.  Comparing \Eq{eq:softmixings} to numerical calculations we find good agreement, with the dominant $\tilde{m}_{23}^2$ component agreeing to within $1 \%$.  The subdominant $\tilde{m}_{13}^2$ component agrees to within $10 \%$, where the increased discrepancy comes from higher orders in the larger mixing angles. The $\tilde{m}_{12}^2$ terms are never relevant.  In this work, we are focussing on the case where the flavor symmetry breaking is dominated by the up-type vevs, hence $v_{d3} \ll v_{u3}$ and the off-diagonal elements in the scalar soft mass matrix are small. As we will discuss further in \Sec{sec:constraints},  much larger off-diagonal elements arise by rotating to the fermion mass eigenbasis, so that the off-diagonal terms in the gauge interaction basis are essentially irrelevant for experimental bounds.  

Revisiting lepton flavor structures, the generation of lepton masses and mixings should arise through vevs of $\SU(3)_F$ charged fields, in analogy with the quarks. However, as long as these vevs are sufficiently subdominant to $\langle \mb{S_u} \rangle$, then the squark and slepton soft masses will remain predominantly determined by the structure of up-type quark Yukawas.  In particular, large mixing angles in the neutrino sector will feed through to off-diagonal elements in the squark and slepton soft mass-squared matrices with a suppression of order $v_{\ell3}^2/v_{u3}^2$.  Thus for the case at hand, it suffices to consider the quark flavor structure alone when considering the scalar soft mass spectrum, and one is not forced to commit to a particular model of neutrino masses.  Of course, one could imagine cases where $v_{\ell 3} \gg v_{u3}$ and the leptonic mass and mixing matrices would be relevant.

\section{Outline of a Complete Model}
\label{sec:completemodel}

\subsection{Gaugino Masses}
\label{sec:gauginos}

Flavor mediation is an elegant way to generate hierarchical soft masses in the squark sector.  Even if messenger fields are uncharged under SM gauge groups, flavor mediation also contributes to SM gaugino masses at the three-loop level, as we show in \App{app:threeloopgaugino}.  While this naturally generates a hierarchy between gauginos and the first-two-generation squarks, the hierarchy is too big, as the gauginos are typically lighter than the third-generation squarks.  Thus, to obtain phenomenologically acceptable gluino masses, we must augment the flavor-mediated soft masses with an additional source of SUSY breaking.

A number of different mediation mechanisms could raise the gaugino masses.  Gaugino mediation \cite{Kaplan:1999ac, Chacko:1999mi} is in some sense the most minimal option, since additional contributions to the stop masses are suppressed relative to gaugino masses, keeping stops within naturalness bounds.  Anomaly mediation or gravity mediation could also be employed, though this would raise the additional question of the coincidence in scales between these soft mass contributions and those from flavor mediation.

Our preferred option is depicted in \Fig{fig:flavgaugediag}, where gauge mediation via SM gauge groups is the additional source of gaugino masses.  From a UV perspective, this situation might even be expected.  If one takes the point of view that all SM gauge groups---including gauged flavor groups---should be treated on an equal footing, then one would expect all gauge groups to transmit SUSY breaking to the SSM.  In this way, the gauged flavor symmetry should really be considered as an additional SM gauge group, which happens to be broken well above the weak scale.  We will explore the spectrum of flavor mediation plus SM gauge mediation in \Sec{sec:exampspectra}.

\subsection{The Higgs Sector}
\label{sec:higgs}

Gauge mediation by itself does not resolve issues in the Higgs sector such as the $\mu$ problem or generating a $B_\mu$ term of the correct size.  On the other hand, because the Higgs bosons are uncharged under $\SU(3)_F$ and $\SU(3)_C$, they do not feel the effects of flavor mediation or color gauge mediation.  Thus, the Higgses are naturally lighter than the squarks, lessening fine tuning tensions in the Higgs sector.  That said, recent hints of a 125 GeV Higgs exacerbate fine tuning tensions in the MSSM, especially in gauge mediation  \cite{Draper:2011hd}.

In our framework, the details of the Higgs sector are largely irrelevant for understanding flavor mediation, but we wish to give an existence proof that a 125 GeV Higgs is compatible with our flavor structure without fine tuning.  This is most easily accomplished in the so-called S-MSSM \cite{Delgado:2010uj}, which is a singlet extension of the MSSM that differs from the NMSSM by the presence of explicit mass terms  for the higgsinos and singlinos
\be\label{eq:smssm}
W_{\rm Higgs} = \mu_H H_u H_d + \mu_S S^2 + \lambda S H_u H_d + f S + \kappa S^3.
\ee
Of course, this superpotential does not address the $\mu$ problem.  It does however alleviate fine tuning because the coupling $\lambda$ generates a quartic coupling which is relevant at small $\tan \beta$, which is also favored by trying to minimize $\alpha$ in \Eq{eq:defalpha}. Such an extension of the Higgs sector can accommodate a Higgs mass in the vicinity of 125 GeV, and may arise dynamically as in \Ref{Craig:2011ev}.  

Because we are agnostic as to the structure of the Higgs sector, we will not be able to give any precise predictions for the masses of electroweak-ino or singlino states.  Following the philosophy of natural SUSY, though, we do expect $\mu_H$ to be in the neighborhood of $200 \GeV$ to minimize cancellations between SUSY and SUSY-breaking Higgs mass contributions.  

\subsection{Example Spectra}
\label{sec:exampspectra}
The addition of SM gauge mediation to flavor mediation introduces a number of new features.  One appealing byproduct is that sfermion mass matrices remain approximately diagonal and the first two generations remain highly degenerate, ameliorating flavor constraints.   Gauge mediation does introduce new representation-dependent splittings, such that sleptons and squarks are no longer degenerate, nor are left- and right-handed sfermions.  Though the third-generation sfermions do receive additional contributions from SM gauge mediation, they remain much lighter than the first two generations.  Since three-loop flavor-mediated gaugino masses are small, the gaugino mass spectrum retains the highly predictive pattern of gauge mediation.

A less obvious feature arises when we consider RG running.  In the simplest examples of gauge-mediated spectra, the stops are typically as heavy as the gluino, and even when RG running drives a stop lighter than the gluino, often only one stop runs lighter and still remains close in mass to the gluino.  This is undesirable for a natural SUSY spectrum, which typically requires the gluino to be heavier than both stops.  When SM gauge mediation is combined with flavor mediation, though, the stops are not only lighter than the first-two-generation squarks at the messenger scale, but RG effects due to the heavy first two generations drive the stop masses \emph{down} at two-loops.  This is in fact the dominant effect that sets an upper bound on the mass of the first-two-generation squarks in natural SUSY models, since if they are too heavy, they drive the stops tachyonic \cite{Dimopoulos:1995mi,Giudice:2008uk}.   In our context, this RG effect typically drives both stops lighter than the gluinos, as desired for natural SUSY.\footnote{One might worry that this RG effect constitutes a hidden fine tuning.  However, the stop masses automatically start off smaller than the first two generations at the messenger scale, and the RG effect only reduces the stop mass from, say, $\mathcal{O}(700)$ GeV to $\mathcal{O}(400)$ GeV.}

Although a complete study of RG behavior in the full parameter space of flavor mediation plus SM gauge mediation is beyond the scope of this work, we will study two benchmark points with an acceptable natural SUSY spectrum of states.  For simplicity, we assume that the same SUSY-breaking spurion $\langle \mb{X} \rangle = M + \theta^2 F$ appears in both flavor mediation and SM gauge mediation, and we fix the ratio $F/M = 100 \TeV$.  We assume that the messengers consist of one vector-like pair of $\SU(3)_{F}$ fundamentals as in \Sec{sec:sfermions}, and one vector-like pair of $\SU(5)_{\rm GUT}$ fundamentals for SM gauge mediation.  Of course one could imagine loosening these assumptions, but it is quite satisfying that the same SUSY-breaking spurion can be used for both kinds of gauge mediation.

\begin{table}[t]
\centering
\begin{tabular}{c | c c c | c c | c c c}
\hline\hline
Benchmark & $M$ [GeV]  & $\sqrt{C (\mb{\Phi})} \alpha_F (M)$ & $\delta$ & $\tilde{m}^F_{1,2}$ [GeV] & $\tilde{m}^F_{3}$ [GeV]  & $m_{\tilde{g}}$ [GeV] & $m_{\tilde{t}_1}$ [GeV] & $m_{\tilde{t}_2}$ [GeV]  \\ [0.5ex] 
\hline
Low Scale & $10^8$  & 0.54 & $129^2$ & 6000 & 300 & 859 & 367 & 575 \\ [0.5ex]
High Scale & $10^{14}$   & 0.32 & $72^2$ & 4000 & 300 & 836 & 332 & 608 \\ [0.5ex]
\hline \hline
\end{tabular}
\caption{Two benchmark models for flavor mediation plus SM gauge mediation. The gauge mediation parameter is set to $F/M = 100$ TeV in both cases, and the messengers consist of one vector-like pair of $\SU(3)_F$ fundamentals and one vector-like pair of $\SU(5)_{\rm GUT}$ fundamentals.  The first five columns show the parameters at the messenger scale $M$, including the $\SU(3)_F$ fine structure constant, flavor-breaking parameter $\delta = g_F^2 v_F^2 /M^2$, and sfermion soft masses $\tilde{m}^F_i$ from flavor mediation alone.  The last three columns show the physical gluino and stop masses after RG evolution to the weak scale, with uncertainties of order 2\%.  The first two generation squarks and sleptons remain close to $\tilde{m}^F_{1,2}$.}
\label{tab:flavgauge}
\end{table}

\begin{figure}[t]
\centering
\includegraphics[height=3.8in]{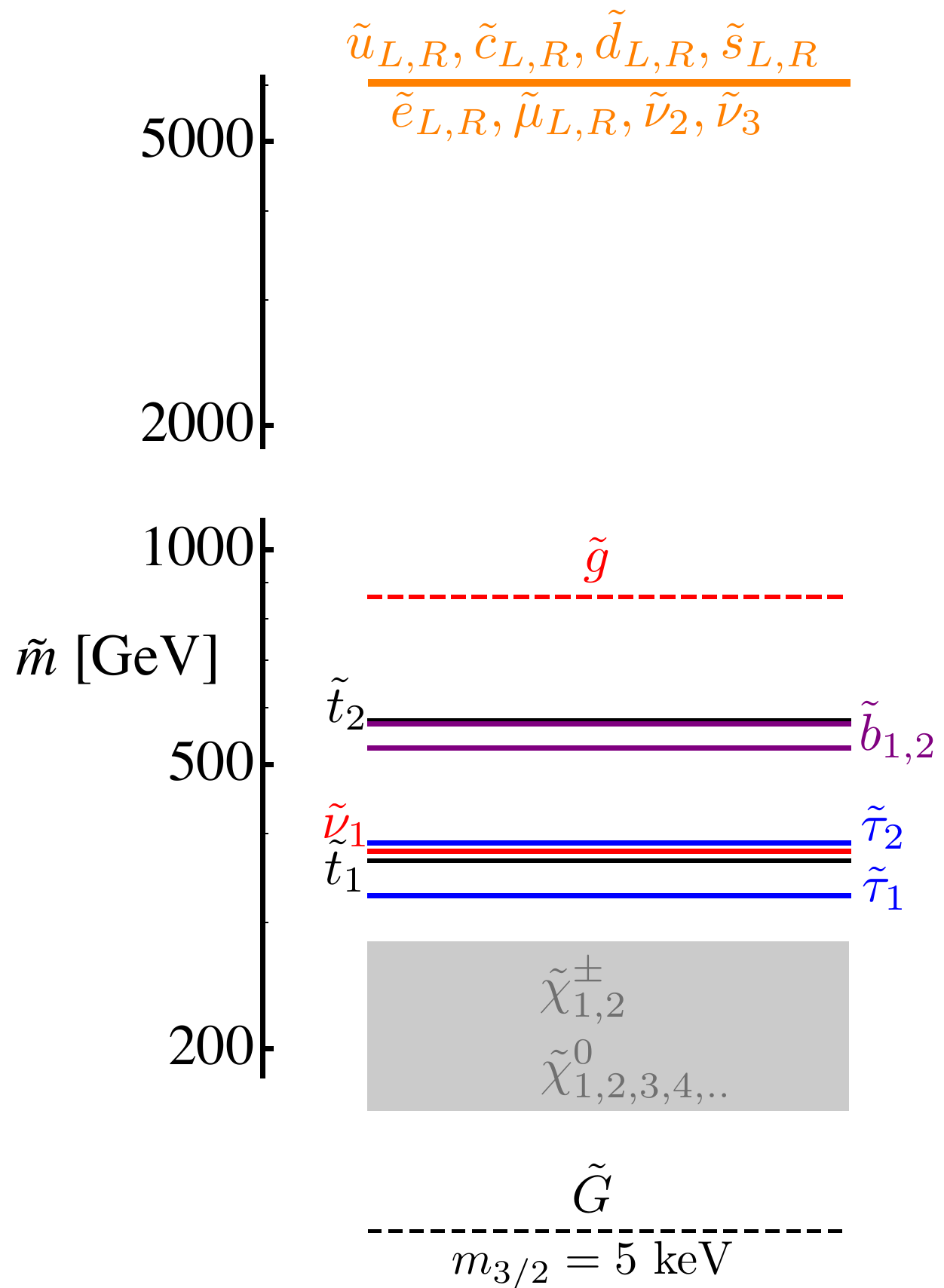}  \hspace{0.75in}  \includegraphics[height=3.8in]{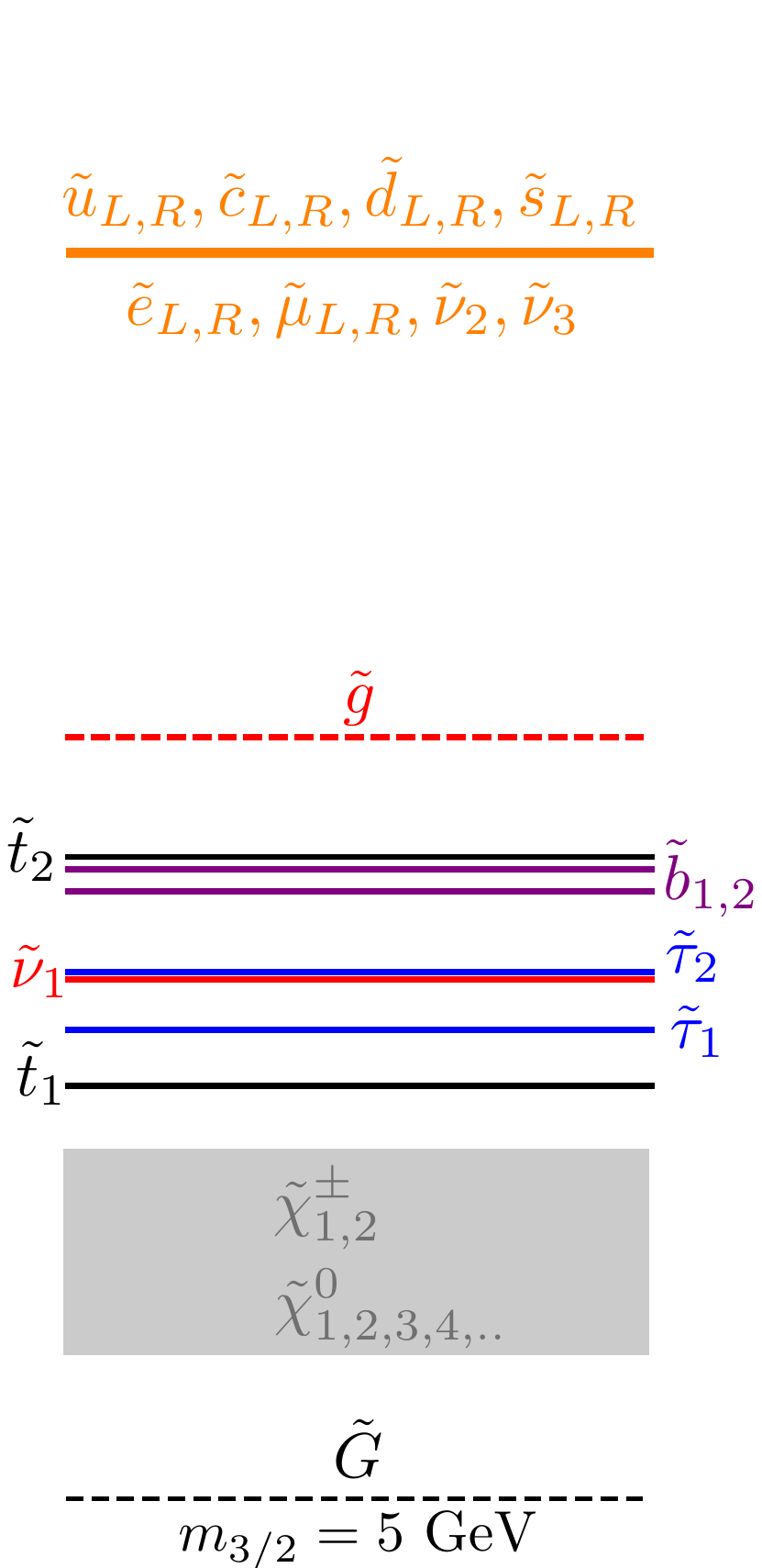}
\caption{Weak scale sparticle spectra for the low (left) and high (right) mediation scale parameters detailed in \Tab{tab:flavgauge}.  Because we are agnostic as to the Higgs sector, we only show the expected mass range for the charginos and neutralinos, guided by naturalness considerations and gauge-mediated expectations.  As expected, the first-two-generation sfermions are largely decoupled, while the third-generation sfermions and gauginos are light.  These spectra are chosen to lie in close proximity to recent collider bounds \cite{Essig:2011qg,Kats:2011qh,Papucci:2011wy,Brust:2011tb}, representing optimistic scenarios for LHC observability.}
\label{fig:spectrum}
\end{figure}

In \Tab{tab:flavgauge}, we show the UV parameters for a low-scale ($M = 10^8 \GeV$) and a high-scale ($M = 10^{14} \GeV$) benchmark.  These benchmarks have been chosen to be representative of the spectra possible within models of flavor mediation.   Both benchmarks are minimal gauge-mediated scenarios with additional contributions from flavor mediation, modeled by adding universal soft masses $\tilde{m}^F_{1,2}$ and $\tilde{m}^F_{3}$ to the first-two- and third-generation sfermions, respectively.  Using SuSpect 2.41 \cite{Djouadi:2002ze}, we RG evolve the UV parameters to achieve the IR spectra depicted in \Fig{fig:spectrum}.  These parameters have been cross-checked with SOFTSUSY 3.3.0 \cite{Allanach:2001kg} and are found to agree within $\mathcal{O}(2\%)$ for most parameters.

The Higgs and electroweak spectra are highly dependent on the specifics of the SUSY Higgs sector, which may involve extra singlets or gauge groups, so we do not specify Higgs sector parameters in order to avoid committing to any specific model.  As the neutralinos and charginos are not charged under the flavor symmetry, we can still estimate their masses.  As stated in \Sec{sec:higgs}, one expects higgsinos should be around $100$--$300$ GeV from naturalness considerations.  The bino and wino obtain masses from gauge-mediated SUSY breaking that gives a characteristic $1:2:6$ mass ratio between the bino, wino, and gluino.  For a $800$--$900$ GeV gluino mass, as is the case for both benchmark models, we expect the bino and wino Majorana masses to also be $100$--$300$ GeV, so it is safe to assume that most (if not all) the electroweak-inos have masses in this range. 

As shown in \Fig{fig:spectrum}, the combination of flavor mediation and SM gauge mediation delivers a natural SUSY spectrum.  The first- and second-generation sfermions do not evolve very much from the messenger scale and remain heavy.  The third-generation sfermions are light because of the suppressed flavor mediation, and exhibit mass splittings from SM gauge mediation between squarks and sleptons as well as between left- and right-handed modes.  For this choice of parameters, the gluino is around a factor of 2 heavier than the lightest stops, which is desirable for natural SUSY and is a result of the RG flow from the messenger scale.

\section{Flavor constraints}
\label{sec:constraints}

Flavor mediation leads to non-universal interactions and soft masses, requiring a careful consideration of constraints from precision flavor measurements. There are two separate sources of flavor-changing neutral currents (FCNCs) in these theories: a set of tree-level contributions coming from the massive $\SU(3)_F$ gauge bosons, and the familiar set of one-loop box diagrams involving squarks and gluinos proportional to the off-diagonal terms in the soft mass matrices. 

We will focus only on quark flavor violation below.  We will not give detailed consideration to leptonic FCNCs, in large part because the leptonic mixing angles are not strongly constrained and depend somewhat on the details of the neutrino sector; thus there are no irreducible limits. However, we note that the leptonic sector enjoys the same $\U(2)$ sflavor symmetry as the quark sector, which guarantees that contributions to the most strongly constrained leptonic FCNCs (such as $\mu \to e \gamma$) will be small even in the presence of large mixing angles in the lepton mass matrix.

\subsection{Tree-Level Flavor Boson Exchange}

Integrating out the massive $\SU(3)_F$ gauge bosons gives rise to a variety of dimension-6 operators of the form
\be
\mathcal{L} \supset - \frac{g_F^2}{2 M_{V_a}^2} (\bar f_M^i \gamma^\mu T^a_{ij} f_M^j)(\bar f_N^k \gamma_\mu T^a_{kl} f_N^l),
\ee
where $M,N = L,R$ and $f$ stands for any (gauge eigenstate) SM fermion.  The $\SU(3)_F$ generators $T^a_{ij}$ are given in the gauge boson mass eigenstate basis, where the masses are given by \Eq{eq:gbmass}, and the generators are labeled in order of decreasing gauge boson masses.  There are no analogous operators involving, e.g., $(h^\dag D_\mu  h)$, since the Higgs multiplets are neutral under $\SU(3)_F$.

 These dimension-6 operators lead to various possible sources of concern. The first comes from flavor-conserving operators that mix fermions of different species, e.g., operators of the form $\bar q \gamma^\mu q \bar \ell \gamma_\mu \ell$. In general, these operators are poorly constrained, and limits require only that $M_{V_a} / g_F \gtrsim$ few TeV. These dimension-6 operators preserve baryon and lepton number, so there is no additional source of proton decay.  Also note that there are none of the usual flavor-conserving operators that contribute to precision electroweak observables, such as $|h^\dag D_\mu h|^2$ and $(h^\dag D_\mu h) \bar q \gamma^\mu q$.

The second (and most important) constraint on tree-level processes comes from flavor-violating operators that contribute to $\Delta F = 2$ FCNC processes. Different $\SU(3)_F$ generators contribute to different FCNC processes.  In the flavor gauge boson mass eigenstate basis, the most important  generators are $|T^2_{13}|$, $|T^3_{23}|$, $|T^4_{13}|$, $|T^5_{23}|$, $|T^7_{12}|$, $|T^8_{12}|$, all of which have a value around 0.5.  The gauge bosons that come from $\SU(3)_F \to \SU(2)_F$ mediate $1 \leftrightarrow 3$ and $2 \leftrightarrow 3$ FCNC processes, while those coming from $\SU(2) \to \emptyset$ mediate $1 \leftrightarrow 2$ processes.  Hence, the lightest gauge bosons introduce contributions to the \emph{most} constrained processes such as $K^0  - \bar K^0$ mixing. 

The strongest constraint arises from the dimension-6 operator 
\be
\frac{1}{\Lambda^2}(\bar d_R^\alpha s_L^\beta)(\bar d_L^\beta s_R^\alpha)
\ee
where Greek indices denote color contractions. The latest limits on this operator require $\Lambda \gtrsim 10^{4}$ TeV assuming no complex phase \cite{Bona:2007vi}. In the presence of an $\mathcal{O}(1)$ CP-violating phase, the bound strengthens to $\Lambda \gtrsim 1.4 \times 10^5$ TeV. In terms of the model parameters here, this implies $v_{u2} \gtrsim 10^4$ TeV ($1.4 \times 10^5$ TeV) without (with) $\mathcal{O}(1)$ CP violation.  Assuming these limits on $v_{u2}$ are satisfied, the rate for all other  $1 \leftrightarrow 3$ and $2 \leftrightarrow 3$ FCNC processes are well below their experimental limits.

\subsection{Gluino-Squark Box Diagrams}
 
Beyond the flavor gauge bosons, the principal constraints arise from off-diagonal sfermion soft masses in the basis where both the fermion masses and the gluino couplings are diagonal.  These off-diagonal sfermion soft masses lead to one-loop contributions to FCNC processes dominated by box diagrams involving squark and gluino exchange. There are two such contributions to these off-diagonal soft mass terms.  One contribution appears in the gauge interaction eigenbasis from $i \neq j$ terms in \Eq{eq:nonab} (whose parametric behavior is given by the second term in \Eq{eq:softmixings}).  The second contribution arises upon going to the fermion mass eigenbasis by diagonalizing the Yukawa textures in \Eq{eq:vevs}.   The gauge interaction eigenbasis contributions are typically much smaller than those contributions coming from diagonalization of the fermion mass matrix, since the latter are suppressed by $v_{d3}^2 / v_{u3}^2$. Thus, to leading order, we may focus on the off-diagonal terms arising solely from rotating to the fermion mass eigenbasis.

For the choice of $\SU(3)_F$ vevs in \Eq{eq:vevs}, the up-type fermion mass matrix is already diagonal in the gauge interaction eigenbasis. The down-type mass matrix may be diagonalized by the transformation $M_d \to V_{\CKM} M_d V_{\CKM}^T$.  SUSY relates this rotation on the fermions to a rotation on the corresponding sfermions.  As such, the up-type scalar mass matrices $\tilde m^2_{u_L}, \tilde m^2_{u_R}$ are unchanged while the down-type scalar mass matrices transform as
\be
 \tilde  m^2_{d_L}  \to V_{\CKM}^\dag \tilde m^2_{d_L} V_{\CKM}, \qquad  \tilde  m^2_{d_R} \to  V_{\CKM}^T\tilde m^2_{d_R} V_{\CKM}^*.
\ee 
In the fermion mass eigenbasis, the diagonal entries in the sfermion mass matrix are of the form
\be
 (\tilde m^2_{d_{L,R}} )_{ii} = |V_{ki}|^2 \tilde m_k^2.  
\ee
The off-diagonal entries are of the form 
\be
(\tilde m^2_{d_L} )_{ij} = V_{2i}^* V_{2j} (\tilde m_2^2 - \tilde m_1^2) + V_{3i}^* V_{3j} (\tilde m_3^2 - \tilde m_1^2) \qquad i \neq j,
\ee
and $(\tilde m^2_{d_R})_{ij} = (\tilde m^2_{d_L})_{ij}^*$.  

The leading contribution to $K^0 - \bar K^0$ mixing arises from a gluino-squark box diagram with two insertions of the off-diagonal soft masses, with additional suppressions coming from the heaviness of first- and second-generation scalars.  However, this contribution is far below current limits, since it is proportional to $(\sin \theta_c \times \tilde \delta_{12})^2 \lesssim 10^{-11}$, where $\theta_c$ is the Cabibbo angle and $\tilde{\delta}_{12} \equiv (\tilde{m}^2_2-\tilde{m}^2_1)/((\tilde{m}^2_2+\tilde{m}^2_1)/2)$ is shown in the right panel of \Fig{fig:sfsplit}. 

A more important contribution arises from a box diagram with {\it four} insertions of the off-diagonal soft masses, at order $(\tilde m_{13}^2 \tilde m_{23}^2)^2$. Although this diagram is suppressed by additional small mixings, the momentum running in the loop diagram is dominated by lighter third-generation squarks. Constraints on such processes were studied in \Ref{Giudice:2008uk}. Here we may directly compute the dominant gluino-mediated contributions to meson mixing appropriate to a $\U(2)$-symmetric hierarchical soft spectrum, which we obtain from \Ref{Bertolini:1990if} using the techniques of \Ref{Giudice:2008uk}. This gives contributions to meson mixing at the scale of the soft masses; limits are computed by RG evolving these contributions to the relevant IR scale as in \Ref{Bona:2007vi}. 

We find that the sbottom-mediated FCNC contributions to the real part of $K^0 - \bar{K}^0$ mixing are an order of magnitude below experimental constraints. Similarly, contributions to the imaginary part of $K^0 - \bar{K}^0$ mixing from the SM CKM phase are below experimental limits. However, if there is an $\mathcal{O}(1)$ new CP-violating phase in the SUSY-breaking soft parameters, the sbottom-mediated contributions to the imaginary part of $K^0 - \bar{K}^0$ mixing would be an order of magnitude larger than allowed by measurements of $\epsilon_K$.  In flavor mediation, if the flavor fields $\mb{S_u}, \mb{S_d}$ do not have significant $F$-term expectation values and the SSM gaugino masses are universal, the only potentially relevant new source of CP violation comes from a relative phase between gaugino masses and the $B_\mu$ parameter. This phase does not enter into the leading gluino-mediated contributions to $\Delta F = 2$ FCNCs, and its contribution to, e.g., electric dipole moments may be rendered negligible if $B_\mu = 0$ at the scale of SUSY breaking. Both real and imaginary contributions to analogous up-type squark processes such as $D^0 - \bar D^0$ mixing are much less tightly constrained and have no large irreducible contributions in flavor mediation.

\begin{figure}[t]
\centering
\includegraphics[height=3.0in]{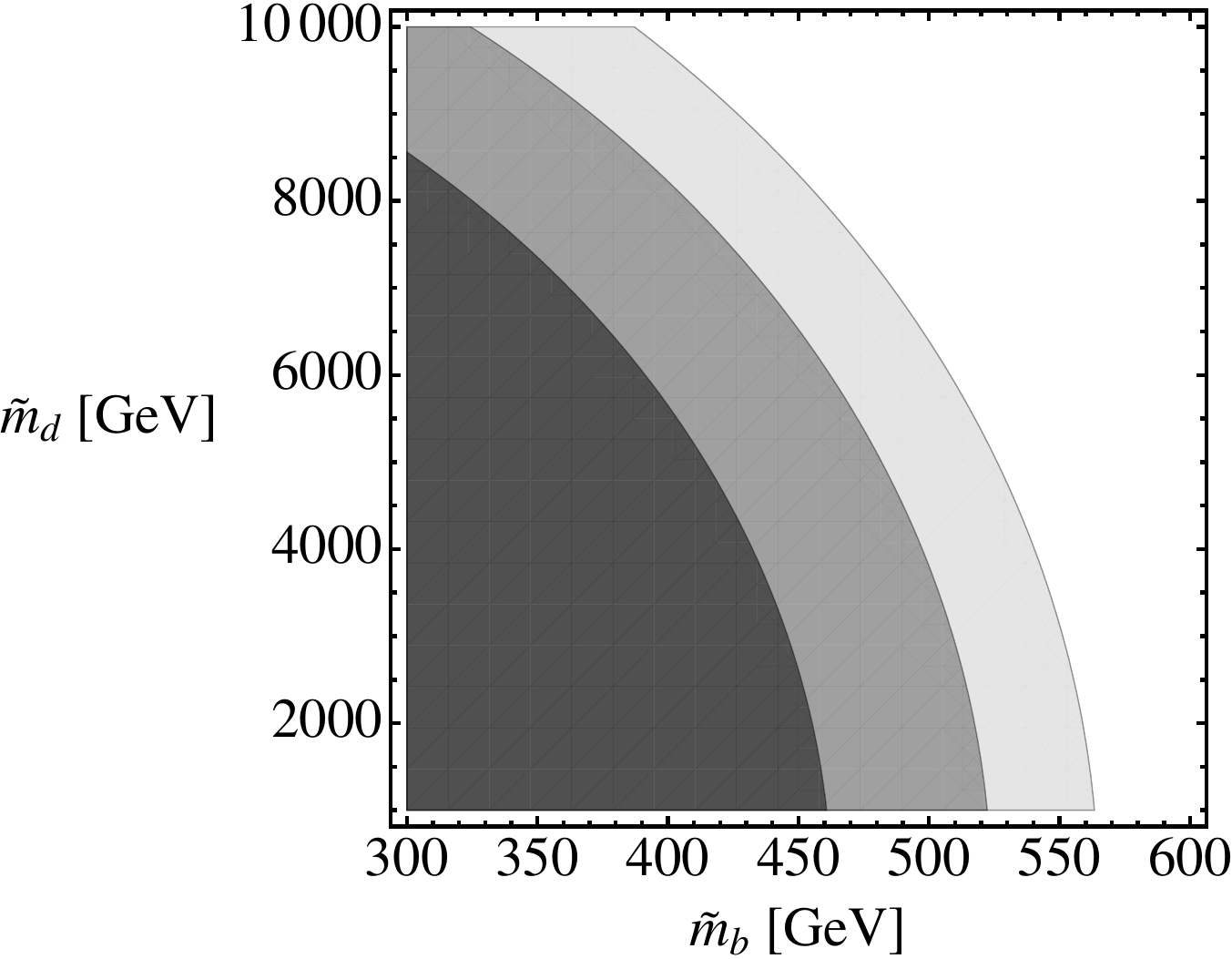}
\caption{Limits on the soft spectrum from the real contribution to $B_d^0 - \bar B_d^0$ mixing, as a function of the average sbottom mass $\tilde m_b$ and the heavy squark mass $\tilde m_d \approx \tilde m_{1,2}$. Other FCNC processes such as (the real part of) $K^0 - \bar K^0$ mixing and the rare decay $B \to X_s \gamma$ place no significant constraint on the soft masses. The gray-shaded exclusion contours correspond to $m_{\tilde g} = 0.7, 1.0$, and $1.3$ TeV from light to dark. The Yukawa textures are given by \Eq{eq:vevs} with $\alpha = 1$.   }
\label{fig:fcnclim}
\end{figure}

For the immediate experimental horizon, the most interesting constraints are from $2 \leftrightarrow 3$ FCNC processes such as $B$-meson mixing, which arise at two insertions of the off-diagonal soft masses and are suppressed only by third-generation scalar masses. We find that contributions from $B_d^0 - \bar B_d^0$ are of the same order as current limits, and remain consistent with experimental limits for  reasonable values of $\tilde m_{b}$ and $m_{\tilde g}$. Representative limits are shown in \Fig{fig:fcnclim}. However, the natural prediction for $2 \leftrightarrow 3$ FCNC processes is within reach of future super-B factories provided $m_{\tilde b} \lesssim 1$ TeV \cite{Aushev:2010bq}. Similar contributions to the rare decay $B \to X_s \gamma$ \cite{Borzumati:1999qt} do not provide significant constraints unless $\tan \beta$ is particularly large. 

Finally, the above discussion assumed $\alpha \simeq 1$.   Had we chosen $\alpha \gtrsim 100$ the gauged flavor symmetry breaking would be driven dominantly by $\mb{S_d}$.  In this case the down-type quarks and squarks would be flavor-aligned and the misalignment would now be between up-type quarks and squarks.  This scenario would also be interesting to consider, as constraints are generally weaker for up-type rather than down-type flavor violation.

\section{Predictions}
\label{sec:predictions}

The degeneracy of the first-two-generation scalars arises as a non-trivial consequence of the mechanism of Higgsed gauge mediation combined with the structure of the SM Yukawas.  Hence, this degeneracy is a key prediction of the model.  This has considerable \ae sthetic appeal when compared to models that impose the degeneracy by hand.   In models that invoke additional $\SU(2)$ or $\U(1)$ symmetries, the degeneracy between the first-two-generation scalars is an input rather than an output of the model.  In flavor mediation, the custodial $\U(2)$ symmetry is tied directly to the peculiar structure of the up-type Yukawa matrix with hierarchical entries.

By requiring that the gauged flavor symmetry is anomaly-free, treats all three flavors equally, and is consistent with GUT structures in the UV, we were led to an $\SU(3)_F$ flavor symmetry under which all matter multiplets are fundamentals.  These restrictions lead to a rather predictive structure for the sfermion soft masses.  In particular, the entire third generation, including leptonic multiplets and the right-handed sbottom, should be energetically accessible at the $8$ TeV LHC.  This is to be contrasted with the most extreme natural SUSY spectrum possible in which the stops and left-handed sbottom are the only flavored scalars within LHC reach.  In our case, the requirement of anomaly cancellation thus leads to the additional prediction of light right-handed sbottom, staus, and a sneutrino.\footnote{One could also imagine a misalignment between the lepton and quark generations such that selectrons or smuons were light instead of the stau, though one would then have to check leptonic FCNCs like $\mu \rightarrow e \gamma$.}

If SM gauge groups also mediate SUSY breaking, then the degeneracy of third-generation scalars is lifted, and the soft masses are dependent on the SM representation.   Majorana gaugino masses follow the predictive $1:2:6$ gauge-mediated pattern.  Gluinos cannot be much more massive than the left-handed stop and should be close to current gluino mass bounds, within immediate LHC reach.  As gaugino masses are generated via gauge mediation, the winos and bino should be close to the weak scale.  As the Higgs multiplets are uncharged under $\SU(3)_F$ and $\SU(3)_C$ they are lighter than the stops.  Consequently, all charginos and neutralinos should be close to the weak scale, and lighter than the stops and gluinos.  To summarize, predictions for the LHC include potential observation of the electroweak sector and gauginos of the SSM, along with the entire third-generation of sfermions.

As in many theories of natural SUSY, contributions to flavor-changing neutral currents are largest among $B$-mesons. While the contributions are consistent with current limits for a wide range of soft masses, they are likely to be measurable at future $B$ factories, particularly via contributions to $B_d^0 - \bar B_d^0$ mixing and the decays $B \to X_s \gamma$.  Moderate new sources of CP violation may also be first apparent in the $B$ meson system while remaining consistent with limits coming from lighter mesons.  If the sbottom and stop are heavy enough that their direct production rate is suppressed, these indirect measurements may provide the first indication of new physics.

Cosmological considerations are similar to those for standard gauge mediation.  One expects the gravitino to be the lightest SUSY particle, and the gravitino phenomenology should remain much the same as in standard gauge mediated scenarios.  For low mediation scales, the gravitino poses no cosmological hazard.  For higher mediation scales, the thermal abundance of the gravitino may be too large, but all of the standard mechanisms to alleviate this challenge in gauge mediation may be employed, such as multiple SUSY breaking \cite{Cheung:2010mc,Cheung:2010qf,McCullough:2010wf} or gravitino decoupling \cite{Luty:2002ff,Craig:2008vs}.

\section{Conclusion}
\label{sec:conclusion}

Independent of the motivations from natural SUSY, flavor mediation is a compelling mechanism to link SUSY flavor to SM flavor.  By gauging the maximum non-Abelian non-anomalous flavor group in the SM (consistent with GUT structures in the UV), we achieve a viable SUSY flavor structure that is neither flavor blind nor minimally flavor violating.  The fact that a custodial $\U(2)$ flavor symmetry is present in these models is a non-trivial feature of $\SU(3)_F$ with large top Yukawa breaking.  The fact that such models are consistent with (and perhaps hinted by) LHC data highlights the importance of flavor structures for physics beyond the SM.

A number of open questions and directions for future investigation remain.  This framework is quite restrictive, with the flavor gauge group and matter superfield representations determined almost entirely by anomaly cancellation, flavor universality, and the desire for gauge coupling unification.  Hence only a few parameters are added in addition to a standard gauge-mediated model, and a full exploration of the parameter space and RG behavior is possible, along with consideration of the residual fine-tunings present in the model. 

In this work, we have remained agnostic about the details of leptonic flavor, which is a reasonable assumption if the size of spontaneous flavor symmetry breaking in the lepton sector is small compared to the quark sector. Nonetheless, it is also possible to construct viable models of flavor mediation with large spontaneous flavor breaking in the lepton sector, which may lead to additional novel features in the slepton spectrum related to neutrino mixing angles.  We have also remained agnostic about the Higgs sector, and it would be interesting to study whether flavor mediation might provide a different perspective on the $\mu$ and $B_\mu$ problems.  Another area relevant to current LHC SUSY searches is the impact of a gauged flavor symmetry on the size and structure of possible $R$-parity-violating operators in the SSM.

Looking towards the UV, it is necessary for the flavor symmetry breaking scale to be close to the messenger masses.  This could be simply a coincidence, or it could be the case that the SUSY-breaking and messenger sectors also determine SM flavor structures.   With this in mind, it would be interesting to embed all of these features within a single dynamical SUSY breaking sector, such as an ISS sector \cite{Intriligator:2006dd}.   More ambitiously, it would be intriguing if all four gauge groups (or corresponding dual descriptions) could be unified into a single group in the UV.  After all, even though the gauged flavor symmetry is broken at scales well above the electroweak scale, it is fundamentally no different from the other SM gauge groups.  

Finally, it may be possible to generalize features of flavor mediation beyond the specific model detailed here.  For example, one could cast the physics in terms of correlation functions for Higgsed gauge mediation \cite{Buican:2009vv}, much as was achieved in the ``general gauge mediation'' program \cite{Meade:2008wd}.  Such studies would highlight the robustness of flavor mediation as a source of SUSY breaking.

\acknowledgements{We thank Asimina Arvanitaki, Edward Hardy, Karoline Kopp, Lisa Randall, Matt Strassler and Giovanni Villadoro for helpful suggestions.  NC is supported in part by the NSF under grant PHY-0907744 and gratefully acknowledges support from the Institute for Advanced Study.  MM and JT are supported by the U.S. Department of Energy (DOE) under cooperative research agreement DE-FG02-05ER-41360.  MM is also supported by a Simons Postdoctoral Fellowship, and JT is also supported by the DOE under the Early Career research program DE-FG02-11ER-41741.}

\appendix

\section{Three-loop Gaugino Masses}
\label{app:threeloopgaugino}

As discussed in \Sec{sec:gauginos}, flavor mediation gives a three-loop contribution to SM gaugino masses.  In this appendix, we calculate this contribution and show why it is too small to yield a realistic natural SUSY spectrum.

The three-loop SM gaugino mass from flavor mediation can be derived using analytic continuation into superspace.  A real gauge coupling superfield $\mb{R}(\mu)$ can be defined in terms of the holomorphic gauge coupling $\mb{S}(\mu)$ and the matter wave function renormalizations $\mb{Z}_r(\mu)$ as \cite{Giudice:1997ni,ArkaniHamed:1998kj}
\be
\mb{R}_\alpha (\mu) = \mb{S}_\alpha(\mu) + \mb{S}_\alpha(\mu)^\dagger - \sum_{\mb{Q}_r} \frac{C_\alpha(\mb{Q}_r)}{8 \pi^2} \log \mb{Z}_r(\mu) + \ldots,
\ee
where $\alpha$ denotes the SM gauge group under consideration, the sum runs over SSM representations $\mb{Q}_r$, and $C_\alpha(\mb{Q}_r)$ is the associated Dynkin index.  The gaugino mass is then given by
\be
m_{\lambda_\alpha} = - g_\alpha^2 \mb{R}_\alpha(\mu)|_{\theta^2}  .
\ee

If there are no messengers charged under the SM, then $\mb{S}(\mu)$ does not contain a $\theta^2$ component, and we need only calculate the $\theta^2$ component of the two-loop $\mb{Z}_r(\mu)$ to extract the three-loop gaugino mass.  For an SSM superfield charged under the mediating gauged flavor symmetry $\SU(3)_F$, we can use the full two-loop effective K\"ahler potential calculated in \Ref{Nibbelink:2005wc,Craig:2012yd}.   Taking appropriate derivatives we have
\be
\mb{Z}_{\Phi_{ij}} (\mu)|_{\theta^2} = C_F(\Phi) \frac{\alpha_F^2}{(2\pi)^2} \frac{F}{M} \sum_a h(\delta^a) (T^a_q T^a_q)_{\{ij\}},
\ee
where $ij$ are (symmetrized) flavor indices, $\alpha_F$ is the flavor group fine structure constant, and $C_F(\Phi)$ is the Dynkin index of the messenger superfields.   The suppression factor is
\be
h(\delta) = 2 \frac{(\delta-4) \delta \log \delta - \Omega(\delta)}{(4-\delta)^2 \delta},
\ee
with $\Omega(\delta)$ defined in \Eq{eq:omega}.  To find the total contribution to gluino soft masses, we must sum over contributions from all SSM quark superfields.  The superfields $\mb{Q},\mb{U^c},\mb{D^c}$ are all in the same representation of $\SU(3)_F$ and hence contribute in the same way to gluino masses. Summing over these fields we generate an additional factor of $4$.  We must also sum over the individual flavors, which contribute differing amounts.  The combined result is
\be
m_{\tilde{g}} = C(\mb{q}) C_F(\mb{\Phi}) \frac{\alpha_S \alpha_F^2}{2 \pi^3} \frac{F}{M}    \sum_i \sum_a h(\delta^a) (T^a_q T^a_q)_{ii} .
\ee

\begin{figure}[t]
\centering
\includegraphics[height=2.8in]{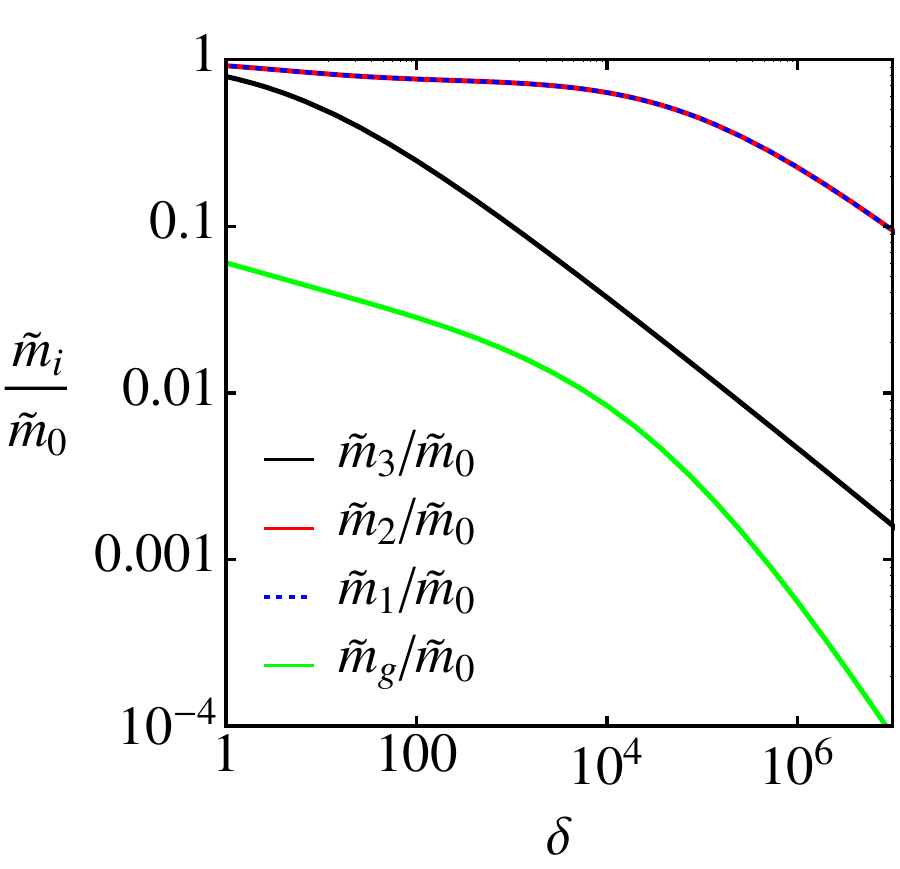}    
\caption{Three-loop gaugino masses in comparison to the sfermion masses as a function of $\delta = g_F^2 v_F^2/M^2$.  The sfermions masses are the same as in \Fig{fig:sfspec}.  For the purposes of demonstration, we have fixed $\alpha_S =0.1$ and we have saturated the perturbativity bound for the flavor group with $\alpha_F = 1$.  The flavor messengers are chosen in the fundamental, such that $C_F(\mb{\Phi}) = 1/2$.  Even under this extreme choice of parameters, the gauginos are typically much less massive than the third-generation sfermions.}
\label{fig:gaugespec}
\end{figure}

The suppression factor $h(\delta^a)$ behaves as
\be
\lim_{\delta \to 0} h(\delta) = (1 - \log \delta), \qquad \lim_{\delta \to \infty} h(\delta) = \frac{2 \log \delta}{\delta} ,
\ee
and hence large logarithms are generated in the limit of small $\delta$.  These logarithms can be thought of as arising due to running effects between the scale of the messenger masses $M$ and the scale at which the flavor symmetry is broken $M_V$.  Whenever
\be
\frac{\alpha_F^2}{2 \pi^2} \log(\delta) \gtrsim 1,
\ee
this signals a breakdown of perturbation theory and the validity of the two-loop calculation.  In this regime, these large logs must be resummed appropriately.  This is a nontrivial exercise, so here we limit $\alpha_F$ and $\delta$ such that these terms are small and perturbation theory still holds.

Back in \Fig{fig:sfspec}, we compared the flavor-mediated squark soft masses to the soft masses that would be generated in the limit of unbroken flavor mediation.  We now compare the three-loop gaugino masses to the same parameter, in order to ascertain whether these three-loop contributions could be great enough to generate an acceptable gaugino spectrum.  The ratio of squared masses is
\be
\frac{m_{\tilde{g}}^2}{\tilde{m}_0^2} = \frac{C(\mb{q})^2 C_F(\mb{\Phi})}{C_2 (\mb{q})} \frac{\alpha_S^2 \alpha_F^2}{\pi^4}  \left(\sum_i  \sum_a h(\delta^a) (T^a_q T^a_q)_{ii} \right)^2.
\label{eq:gauginom0}
\ee
Once the flavor breaking parameters $\delta^a$ have been chosen in order to generate a particular squark soft mass hierarchy, the relative soft masses of the gauginos can only be adjusted by varying $\alpha_F$ and the matter representations.  In \Fig{fig:gaugespec} we show the relative scales of the three-loop gluino masses in comparison to the sfermion masses, choosing the most optimistic value $\alpha_F = 1$.  

Even by maximizing the three-loop contributions from flavor mediation, it would be difficult to achieve the desired gluino and stop masses for a natural SUSY spectrum using flavor mediation alone.  One could choose larger representations of flavor messengers to raise the relative gaugino masses, however it is clear that the three-loop contributions to gaugino masses alone are generally not sufficient to allow for a natural SUSY spectrum with $m_{\tilde{t}} \lesssim 400$ GeV and $m_{\tilde{g}} \gtrsim 600$ GeV.  In \Sec{sec:gauginos}, we therefore introduced an additional contribution from SM messengers.  Of course, an alternative strategy would be to introduce messengers charged under both $\SU(3)_F$ and the SM gauge group.

\bibliography{FLAVMEDref}

\end{document}